\documentstyle[12pt,a4]{article}

\newcommand{\be}{\begin{equation}}
\newcommand{\ee}{\end{equation}}
\newcommand{\beq}{\begin{eqnarray}}
\newcommand{\eeq}{\end{eqnarray}}
\newcommand{\AmS}{{\protect\the\textfont2
  A\kern-.1667em\lower.5ex\hbox{M}\kern-.125emS}}

\begin{document}
\begin{flushright}
{TIFR/TH/99-37}\\
{July, 1999}
\end{flushright}

\vskip 2em

\begin{center}
{A Minimal See-Saw Model for Hierarchical Neutrino Masses with
Large Mixing\footnote{To appear in the proceedings of the 6th
San Miniato Topical Seminar on `Neutrino and Astroparticle Physics',
Nuclear Physics B (Proc. Suppl.)}} \\ [4mm]
{D.P. Roy} \\  
{Tata Institute of Fundamental Research, Mumbai 400 005, India}
\end{center}

\begin{abstract}
The atmospheric and solar neutrino oscillation data suggest
hierarchical neutrino masses with at least one large mixing. The
simplest see-saw models for reconciling the two features are $U(1)$
extensions of the SM with flavour dependent gauge charges. I discuss
a minimal model of this type containing two heavy right-handed
neutrinos, which have normal Dirac couplings to $\nu_\mu$ and
$\nu_\tau$ but suppressed ones to $\nu_e$. It can naturally account for
the large (small) mixing solutions to the atmospheric (solar) neutrino 
oscillation data.
\end{abstract}



The recent Superkamiokande data has provided convincing evidence for
atmospheric neutrino oscillation and confirmed earlier results of
solar neutrino oscillation \cite{fuku98}. The atmospheric neutrino
data seems to imply a large mixing between $\nu_\mu$ and $\nu_\tau$,
$\sin^2 2 \theta_{\mu\tau} > 0.86$, along with $\Delta M^2 = (1.5 - 6) 
\times 10^{-3}$ eV$^2$ at 90\% CL. They correspond to 
\be
\theta_{\mu\tau} = 45 \pm 11^\circ 
\ee
and
\be
\Delta M \simeq 0.06 \ eV ,
\ee
the latter representing the central value of $\Delta M$ for
hierarchical masses and an upper limit on this quantity for degenerate 
ones. By far the simplest explanation of the solar neutrino
oscillation data is provided by the small mixing angle MSW solution
although one can get equally good descriptions in terms of the large
mixing angle MSW or vacuum oscillation solutions. The SMA solution
corresponds to a small mixing of $\nu_e$ with one of the above states, 
$\sin^2 2 \theta_e = 10^{-3} - 10^{-2}$, along with a small $\Delta
m^2 = (0.5 - 1) \times 10^{-5}$ eV$^2$. They correspond to
\be
\sin\theta_e = (1-5) \times 10^{-2} 
\ee
and 
\be
\Delta m \simeq 0.003 \ eV . 
\ee

By itself the atmospheric neutrino oscillation result of Eqs. (1,2)
could be naturally explained in terms of a nearly degenerate pair of
$\nu_\mu$ and $\nu_\tau$. Indeed a pseudo-Dirac mass matrix for this
pair would lead to degenerate masses and maximal mixing on 
diagonalisation, i.e.
\be
\pmatrix{0 & M \cr M & 0} \rightarrow \pmatrix{M & 0 \cr 0 & - M} , 
\ \ \ \theta = 45^\circ .
\ee
But explaining the solar neutrino oscillation result of Eqs. (3,4)
would then imply an even finer level of degeneracy between $\nu_e$ and 
one of this pair, which is totally ad-hoc. Therefore it is generally
considered more natural to interpret them as hierarchical states,
i.e.,
\beq
m_1 \simeq \Delta M &\simeq& 0.06 \ eV , \nonumber \\ [2mm]
m_2 \simeq \Delta m &\simeq& 0.003 \ eV , \nonumber \\ [2mm]
m_3 \ll m_2 &\simeq& 0 ,
\eeq
where the first two states are large admixtures of $\nu_\mu$ and
$\nu_\tau$ and the third one is dominantly $\nu_e$. Indeed much of the 
recent literature on neutrino physics is focussed on theoretical
models, mainly in the see-saw frame work, which can naturally
reconcile such hierarchical masses with large mixing
\cite{alta99}. Note that the mass of the 3rd state can be exactly zero 
as far as the atmospheric and solar neutrino oscillation data are
concerned. Thus a minimal see-saw model for explaining these
oscillations requires two right-handed neutrinos with normal Dirac
couplings to $\nu_\mu$ and $\nu_\tau$, but suppressed ones to $\nu_e$.

It may be noted here that the standard see-saw model \cite{gell79}
represents a $U(1)$ extension of the standard model (SM) gauge group
into 
\be
SU(3)_C \times SU(2) \times U(1)_Y \times U(1)_{Y'}
\ee
with the gauge charge \cite{mars80}
\be
Y' = B - L = B - (L_e + L_\mu + L_\tau) .
\ee
Then the requirement of anomaly cancellation implies the existence of
three right-handed singlet neutrinos ($N_i$) with $Y' = -1$ to match
the three left-handed neutrinos ($\nu_{e, \mu, \tau}$) carrying this
gauge charge. Cancellation of the axial parts of the $Y'$ current
between the left and right handed fermions ensures purely vector
coupling for $Y'$, which in turn ensures that the model is anomaly
free \cite{ma98}. The flavour independence of $Y'$ implies however
that the singlet neutrinos have normal Dirac couplings to all the
left-handed doublets $\nu_{e, \mu, \tau}$ along with the SM Higgs
doublet $\phi$ instead of preferential couplings to $\nu_{\mu, \tau}$ 
as suggested by data. In order to accomplish the latter one has to
invoke a horizontal symmetry with flavour dependent charges
\cite{alta99}. In other words one first takes a flavour blind step
beyond the SM and then applies correctives via additional symmetry
groups with flavour dependent charges. Let us consider instead a
one-step process, where the desired flavour depence is incorporated
into the gauge charge $Y'$ of the $U(1)$ extension of SM
(Eq. 7). While such flavour dependent $U(1)$ extensions of the SM
gauge group are hard to embed in the familiar GUTs they can arise
naturally from string theories \cite{iban94}.

We have studied two such $U(1)$ extensions of the SM
\cite{madp98,madp99}, corresponding to the gauge charges 
\be
Y' = B - 3 L_e 
\ee
and
\be
Y' = B - {3\over 2} (L_\mu + L_\tau) ,
\ee
in the context of the atmospheric and solar neutrino oscillations. I
shall concentrate on the simpler of the two models \cite{madp99},
corresponding to the gauge charge (10). Indeed it seems to represent a 
minimal see-saw model for explaining these neutrino oscillation
data. In this case the anomaly cancellation requirement implies the
existence of two right-handed singlet neutrinos $(N_{1,2})$ with $Y' = 
- {3\over 2}$ to match the two left-handed neutrinos $(\nu_{\mu,
\tau})$ carrying this gauge charge.

The minimal Higgs sector of this model consists of
\be
\pmatrix{\phi^+ \cr \phi_0}_{Y'=0} \ \ \ \ \& \ \ \ \ \chi^0_{Y'=-3} ,
\ee
i.e. the SM Higgs doublet along with a singlet carrying non-zero $Y'$
charge. The $Y'$ symmetry is spontaneously broken via the vacuum
expectation value of $\chi$, $<\chi>$, at a high mass scale. The
coupling of this $\chi$ to $\bar N^C_1 N_1$ and $\bar N^C_2 N_2$
gives them large Majorana masses $\sim <\chi>$. Moreover the
coupling of $\phi$ to $\bar\nu_\mu N_{1,2}$ and $\bar\nu_\tau N_{1,2}$
gives them Dirac masses $\sim <\phi>$, while there is no such
coupling to $\nu_e$. Thus the see-saw mechanism would generate two
non-zero mass states, which are large admixtures of $\nu_\mu$ and
$\nu_\tau$, while $\nu_e$ remains massless. 

One can generate a small mixing of $\nu_e$ with the non-zero mass
states, as required by the SMA solution (3) to the solar neutrino
oscillation, by expanding the Higgs sector. For this purpose we add
another doublet and a singlet with
\be
\pmatrix{\eta^+ \cr \eta^0}_{Y'=-3/2} \ \ \ \ \& \ \ \ \
\zeta^0_{Y'=-3/2} .
\ee
The coupling of the doublet $\eta$ to $\bar\nu_e N_{1,2}$ generates
Dirac mass terms $\sim <\eta>$. The singlet $\zeta^0$ does not
couple to fermions; but it is required to avoid an unwanted Goldstone
boson. The latter comes about because there are 3 global $U(1)$
symmetries, corresponding to rotating the phases of $\phi$, $\eta$ and 
$\chi^0$ independently in the Higgs potential, while only 2 local
$U(1)$ symmetries are spontaneously broken. The addition of the
singlet $\zeta^0$ introduces two more terms in the Higgs potential,
$\eta^+ \phi \zeta^0$ and $\chi^0 \zeta^0 \zeta^0$, so that the phases 
can no longer be rotated independently. While the $\zeta^0$ is
expected to acquire a large vev at the $U(1)_{Y'}$ symmetry breaking
scale, the doublet $\eta$ must have a positive mass squared term in
order to avoid $SU(2)$ breaking at this scale. Nonetheless it can
acquire a small but non-zero vev at the $SU(2)$ symmetry breaking
scale, which can be estimated from the relevant part of the potential 
\be
m^2_\eta \eta^\dagger \eta + \lambda (\eta^\dagger \eta) (\chi^\dagger \chi) +
\lambda^\prime (\eta^\dagger \eta) (\zeta^\dagger \zeta) - \mu
\eta^\dagger \phi \zeta . 
\ee
Although we start with a positive $m^2_\eta$ term, after minimization
of the potential with respect to $\eta$ we see that this field has
acquired a small vev,
\be
<\eta> = \mu <\phi> <\zeta> / 2M^2_\eta ,
\ee
where $M^2_\eta = m^2_\eta + \lambda <\chi>^2 + \lambda^\prime
<\zeta>^2$ represents the physical mass of $\eta$. The size of the 
soft term is bounded by the $Y'$ symmetry breaking scale, i.e. $\mu
\leq <\zeta>$. Thus with a choice of $M_\eta \sim 5 <\zeta>$, we get
\be
<\eta> / <\phi> \sim 1/50 ,
\ee
which will account for the small mixing angle of $\nu_e$ (3).

Let us write down the $5\times 5$ neutrino mass-matrix in this
model. We shall be working in the basis where the charged lepton mass
matrix, arising from their couplings to the SM Higgs boson $\phi$, is
diagonal. This defines the flavour basis of the doublet
neutrinos. Since the two singlet neutrinos do not couple to the
charged leptons, their Majorana mass matrix can be independently
diagonalised in this basis. While the overall size of their masses
will be at the $Y'$ symmetry breaking scale, it is reasonable to
assume a modest hierarchy between them,
\be
M_1 / M_2 \sim 1/20 ,
\ee
in analogy with those observed in the quark and the charged lepton
sectors. This will account for the desired mass ratio for the doublet
neutrinos (6). Thus we have the following $5\times 5$ mass matrix
${\cal M}$ in the basis $(\nu_e , \nu_\mu, \nu_\tau, N^C_1, N^C_2)$:
\be
\pmatrix{0 & 0 & 0 & f^1_e <\eta> & f^2_e <\eta> \cr 
         0 & 0 & 0 & f^1_\mu <\phi> & f^2_\mu <\phi> \cr
         0 & 0 & 0 & f^1_\tau <\phi> & f^2_\tau <\phi> \cr 
         f^1_e<\eta> & f^1_\mu <\phi> & f^1_\tau <\phi> & M_1 & 0 \cr
         f^2_e<\eta> & f^2_\mu <\phi> & f^2_\tau <\phi> & 0 & M_2} ,
\ee
where the $f^{1,2}_{e,\mu\tau}$ are the Higgs Yukawa couplings. We
shall assume these couplings to be of similar order of magnitude,
i.e. the elements of a mass-matrix arising from the same Higgs vev are 
expected to be of similar size. There is of course no conflict between 
such democratic mass-matrix elements and the hierarchical mass
eigen-values assumed above (16). In fact they are closely related -
the former implies large cancellation in the determinant as required
by the latter. 

The resulting $3\times 3$ mass-matrix for the doublet neutrinos is
given by the see-saw formula in this basis,
\be
m_{ij} = {D_{1i} D_{1j} \over M_1} + {D_{2i} D_{2j} \over M_2} , 
\ee
where $D$ is the $2\times 3$ Dirac mass matrix at the bottom left of
(17). One can then calculate the corresponding mass eigen-values
$m_{1,2,3}$ and mixing-angles by diagonalising this matrix
\cite{madp99}. Alternatively we can read off the approximate
magnitudes of these quantities directly from the mass matrix (17),
i.e. 
\beq
\tan\theta_{\mu\tau} &\simeq& {\cal M}_{42} / {\cal M}_{43} \simeq
f^1_\mu / f^1_\tau \sim 1 , \nonumber \\ [2mm] 
\sin\theta_e &\simeq& {\cal M}_{51} / {\cal M}_{52} \simeq <\eta> / <\phi> \sim
{1\over 50} , \nonumber \\ [2mm]
m_2/m_1 &\simeq& M_1/M_2 \sim 1/20 .
\eeq
They are clearly in good agreement with the corresponding experimental 
quantities of Eqs. (1), (3) and (6). Note that in this model the
$\nu_e$ mixing with the higher mass $(m_1)$ eigen-state is also
expected to be of similar size as above, i.e.,
\be
{\cal M}_{41} / {\cal M}_{42} \simeq <\eta>/<\phi> \sim 1/50 .
\ee
This prediction is well within the present experimental limit on this
quantity $(\leq 0.2)$ from CHOOZ data \cite{appo98}; but can be tested 
by future long base line experiments.

Finally, the scale of the $Y'$ symmetry breaking can be estimated from 
the larger Majorana mass $M_2$, i.e.
\be
M_2 \sim f^2 <\phi>^2/m_2 \sim f^2 10^{16} {\rm GeV} \sim 10^{12-16} 
{\rm GeV} .
\ee
The lower limit corresponds to $f \sim 10^{-2}$ as in the case of
$\tau$ Yukawa coupling, while the upper limit corresponds to $f \sim
1$ as in the case of top. Thus the observed scale of neutrino masses
(6) can be explained if one assumes the $Y'$ symmetry breaking scale
to be in the range of $10^{12} - 10^{16}$ GeV. 

One can get a more exact derivation of the masses and mixing angles
via the $3\times 3$ mass-matrix of the doublet neutrinos (18), i.e.
\be
\pmatrix{c^2_1+c^2_2 & c_1a_1+c_2a_2 & c_1b_1+c_2b_2 \cr c_1a_1+c_2a_2 
& a^2_1+a^2_2 & a_1b_1+a_2b_2 \cr c_1b_1+c_2b_2 & a_1b_1+a_2b_2 &
b^2_1+b^2_2} , 
\ee
where
\be
a_{1,2} = {f^{1,2}_\mu <\phi> \over \sqrt{M_{1,2}}} , \ \ \ \  
b_{1,2} = {f^{1,2}_\tau <\phi> \over \sqrt{M_{1,2}}} , \ \ \ \  
c_{1,2} = {f^{1,2}_e <\eta> \over \sqrt{M_{1,2}}} .
\ee
Note that the assumed hierarchies of (15) and (16) imply 
\be
a_1, b_1 \gg a_2, b_2, c_1 \gg c_2 .
\ee
The determinant of (22) vanishes identically, ensuring that one of the 
mass eigenvalues is zero. The other two are 
\be
m_1 \simeq a^2_1 + b^2_1 , \ \ m_2 \simeq {(a_1b_2 - a_2b)^2 \over
a^2_1 + b^2_1} .
\ee
The corresponding mixing matrix $U$ between the flavour and the mass
eigenstates is
\be
\pmatrix{\nu_e \cr \nu_\mu \cr \nu_\tau} = \pmatrix{1 & {-c_2\sqrt{a^2_1+b^2_1} \over a_1b_2-b_1a_2} & {c_1 \over 
\sqrt{a^2_1+b^2_1}} \cr 
{b_1c_2-c_1b_2 \over a_1b_2-b_1a_2} & {b_1 \over \sqrt{a^2_1+b^2_1}} & 
{a_1 \over \sqrt{a^2_1+b^2_1}} \cr
{c_1a_2-a_1c_2 \over a_1b_2-b_1a_2} & {-a_1 \over \sqrt{a^2_1+b^2_1}}
& {b_1 \over \sqrt{a^2_1+b^2_1}}} \pmatrix{\nu_3 \cr \nu_2 \cr \nu_1} .
\ee
One can easily check that the Eqs. (23-26) lead to the masses and
mixing angles of Eqs. (19-21).

\end{document}